\documentclass[a4paper]{jpconf}
\usepackage{graphicx}
\usepackage{amsmath}

\def\om{\omega}
\def\pt{\partial}

\def\mev{{\rm MeV}}
\begin{document}
	
\title{The effect of inclusion of $\Delta$ resonances in relativistic mean-field model with scaled hadron masses and coupling constants}

\author{K A Maslov$^1$,  E E Kolomeitsev$^2$, D N Voskresensky$^1$}

\address{$^1$National Research Nuclear University MEPhI (Moscow Engineering Physics Institute), Kashirskoe highway 31, 115409 Moscow, Russia}
\address{$^2$Matej Bel University, Tajovsk\'eho 40, 97401, Bansk\'{a} Bystrica, Slovakia}

\ead{maslov@theor.mephi.ru}
\begin{abstract}
Knowledge of the equation of state of the baryon matter plays a decisive role in the description of neutron stars. With an increase of the baryon density the filling of Fermi seas of hyperons and $\Delta$ isobars becomes possible. Their inclusion  into standard relativistic mean-field models results in a strong softening of the equation of state and a lowering of the maximum neutron star mass below the measured values. We extend a relativistic mean-field model with scaled hadron masses and coupling constants developed in our previous works and take into account now not only hyperons but also the $\Delta$ isobars. We analyze available empirical information to put constraints on coupling constants of $\Delta$s to mesonic mean fields. We show that the resulting equation of state satisfies majority of presently known experimental constraints.
\end{abstract}

\section{Introduction}
Relativistic mean-field  (RMF) models are widely used to construct a realistic equation of state (EoS), which could satisfy various experimental constraints~\cite{Klahn:2006ir}. The most challenging task is to reconcile the constraint on the pressure in the isospin-symmetrical matter (ISM), the so-called "flow constraint"~\cite{Danielewicz:2002pu}, which favors a soft EoS, and the existence of the neutron star (NS), with the mass $M = 2.01 \pm 0.04 \, M_\odot$~\cite{Antoniadis:2013pzd} ($M_\odot$ is the Sun mass), which favors a stiff EoS. In standard RMF models hyperons may appear in NS cores. This results in a decrease of the maximum NS mass below the observed limit, cf.~\cite{Vidana:2015rsa} and references therein.  The appearance of $\Delta$ isobars in NSs would lead to a  further decrease of the maximum NS mass~\cite{Drago2014}. These problems are dubbed in the literature as the hyperon and $\Delta$ puzzles. In~\cite{Maslov:2015msa,Maslov:2015wba} we proposed an RMF model with scaled hadron masses and coupling constants (labeled there as MKVOR), which solves the hyperon puzzle and fulfills successfully  the maximum mass constraint, the flow constraint and various other constraints, but the $\Delta$s were not included. Here we show that our model passes the constraints also with the inclusion of $\Delta$s. A more detailed consideration can be found in~\cite{Kolomeitsev:2016ptu}.
	
\section{The RMF model with scaled hadron masses and couplings}
The Lagrangian of the model is formulated in \cite{Maslov:2015msa, Maslov:2015wba}. The model is a generalization of the non-linear Walecka model with effective coupling constants $g_{mb}^{*} = g_{mb} \chi_{mb}(\sigma)$ and hadron masses $m_i^{*} = m_i \Phi_i(\sigma)$ dependent on the $\sigma$ field \cite{Kolomeitsev:2004ff}, here $m = \{\sigma, \omega, \rho, \phi\}$ lists the included mesonic fields, $b=(N,H,\Delta)$  indicates the baryon species ($N=p,n$; $H$ is the $\Lambda,\Sigma,\Xi$ hyperon; $\Delta$ labels the $\Delta$ isobar) and $i=(b,m)$ labels the hadrons; $\chi_{mb}(\sigma)$ and $\Phi_i(\sigma)$ are the corresponding scaling functions. In the RMF approximation the contribution of $\Delta$s to the energy density has the same form as for spin-$1/2$ fermions but with the spin degeneracy factor  $4$. The energy density  of our model,
\begin{align}
&E[{\{n_b\}},\{n_l\},  f] =  \sum_b E_{\rm kin}\big(p_{{\rm F},b}, m_b^*
(f), s_b \big) +
\sum_{l=e,\mu} E_{\rm kin}(p_{{\rm F},l},m_l, s_l) +\frac{m_N^4 f^2}{2 C_\sigma^2 } \eta_\sigma(f) \nonumber \\ &
+\frac{1}{2 m_N^2} \Big[\frac{C_\om^2 n^2}{ \eta_\om(f)}
+\frac{C_\rho^2 n_I^2}{\eta_\rho(f)}
+ \frac{C_\phi^2 n_S^2}{\eta_\phi(f)}
\Big] , \,
n=\sum_b x_{\om b} n_b,\,
n_I=\sum_b x_{\rho b} t_{3 b} n_b,\,
n_\phi=\sum_{H} x_{\phi H} n_H,
\label{edensity}
\end{align}
is expressed in terms of the scalar field $f = {g_{\sigma N}^{*}(\sigma) \sigma}/{m_N}$, $x_{mb}=g_{mb}/g_{mN}$ and particle densities $n_j=(2s_j+1)p_{{\rm F},j}^3/6\pi^2$, where $j=(b,l)$, $s_j$ is the fermion spin and $t_{3b}$ stands for the isospin projection of baryon $b$. The fermion kinetic energy density is defined as
$
E_{\rm kin}(p_{{\rm F}},m,s) = (2s + 1)\int_0^{p_{{\rm F}}}
\frac{p^2 dp}{2 \pi^2} \sqrt{p^2 + m^2}
$\,.
Meson coupling constants, masses and  scaling functions enter the energy density only in combinations
\begin{gather}
C_M = {g_M m_N}/{m_M}, \quad C_\phi = {g_\om m_N}/{m_\phi}, \quad \eta_m = {\Phi^2_m(f)}/{\chi^2_m(f)}, \quad M = \sigma, \om, \rho\,.
\end{gather}
Therefore, studying infinite matter we do not have to specify the coupling constants, meson masses and their scaling factors separately. The baryon mass scaling function is $\Phi_b = 1 - x_{\sigma b} ({m_N}/{m_b}) f$.  The particle densities as functions of the total baryon density $n$ in beta-equilibrium matter (BEM) follow from the conditions: $\mu_b = \mu_n - Q_b \mu_l$, where $Q_b$ is the baryon charge  and $\mu_j = \frac{\pt E}{\pt n_j}$ are the fermion chemical potentials; $\mu_e=\mu_\mu$; and the charge neutrality condition $\sum_b Q_b n_b - n_e - n_\mu = 0$. These equations are solved self-consistently with the equation of motion for the scalar field $\pt E/ \pt f = 0$. Finally, the pressure is given by $P = \sum_{j } \mu_j n_j - E$.

The parameters of the model in the nucleon sector are fitted to reproduce the nuclear matter saturation properties, which are defined as coefficients of the Taylor expansion of the energy per particle in ISM in terms of $\epsilon = (n - n_0)/3 n_0$ and $\beta = (n_n - n_p) / n$,
$ \mathcal{E} = \mathcal{E}_0 +\frac{K}{2}\epsilon^2
-\frac{K^{'}}{6}\epsilon^3 +\beta^2\widetilde{J}(n) +\dots$ and $\widetilde{J}(n)=\widetilde{J}+
L\epsilon +\frac{K_{\rm sym}}{2}\epsilon^2+\dots
$\,.
We consider an extension of the MKVOR model introduced in~\cite{Maslov:2015msa,Maslov:2015wba}, labeled as MKVOR*, for details see~\cite{Kolomeitsev:2016ptu}. The input parameters are the same as in the MKVOR model: the nuclear saturation density $n_0 = 0.16 \, {\rm fm}^{-3}$, the binding energy ${\cal E}_0 = -16 \, {\rm MeV}$, the incompressibility $K = 250\,{\rm MeV}$, the symmetry energy $\widetilde J = 32\,{\rm MeV}$ and the nucleon effective mass  $m_N^*(n_0) = 0.73 \, m_N$.

\begin{figure}
\centering
\includegraphics[width= \textwidth]{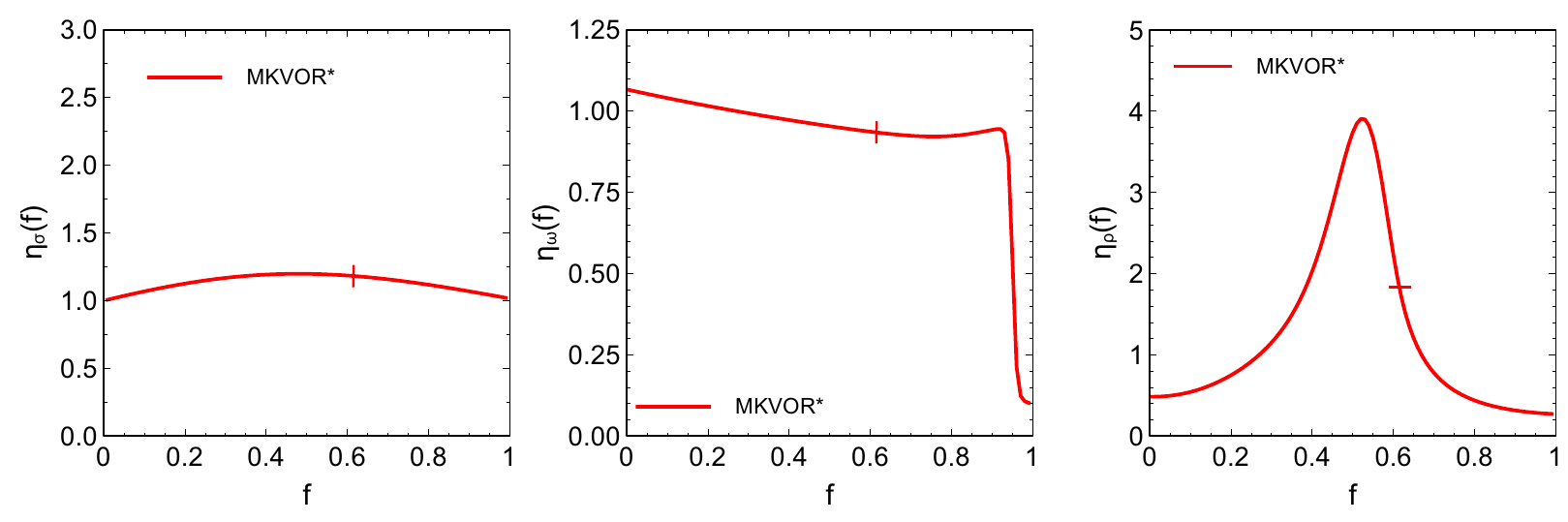}
\caption{Scaling functions $\eta_\sigma$ (left panel), $\eta_\omega$ (middle panel) and $\eta_\rho$ (right panel) for the MKVOR* model. Dashes indicate maximum  values of  $f(n)$ reachable in NSs.
}\label{eta_f}
\end{figure}

Scaling functions of the MKVOR* model as functions of the scalar field are shown in figure~\ref{eta_f}. Compared to MKVOR model in the $\eta_\om(f)$ function an additional sharp decrease at $f > 0.95$ is introduced. A sharp decrease of at least one of $\eta_M (f)$ scaling functions for $f>f_c$ leads to a stiffening of the EoS, cf.~\cite{Maslov:cut}. It limits the scalar field growth with the increasing density and prevents the nucleon effective mass from vanishing, which would occur in the original MKVOR model in the ISM if $\Delta$ were included. Dashes in figure~\ref{eta_f} mark maximum values of the scalar field $f_{\rm max}$, reachable in the  NSs. In the BEM all results for MKVOR and MKVOR* models coincide for densities reachable in NSs.  The behavior of the scaling functions for $f > f_{\rm max}$ is irrelevant in BEM.

The hyperon and $\Delta$ couplings with vector mesons are related to the nucleon ones via the SU(6) symmetry relations~\cite{vanDalen:2014mqa}:
\begin{gather}
g_{\om \Lambda} =g_{\om \Sigma}= 2g_{\om \Xi} = {\textstyle\frac{2}{3} } g_{\om N}, \quad
g_{\rho \Sigma} = 2 g_{\rho \Xi} = 2 g_{\rho N},  \quad g_{\rho\Lambda}=g_{\phi N} = 0 \,,
\nonumber\\
2 g_{\phi \Lambda} = 2 g_{\phi \Sigma} = g_{\phi \Xi} = - {\textstyle\frac{2\sqrt{2}}{3} } g_{\omega N}, \quad
	g_{\omega \Delta} = g_{\omega N}, \quad g_{\rho \Delta} = g_{\rho N}, \quad g_{\phi \Delta} = 0\,.
\end{gather}
Coupling constants with the $\sigma$ meson follow from the baryon potentials in ISM at the saturation density $U_{b}= C_{\omega}^2 m_N^{-2}x_{\omega b}n_0
- (m_N-m_N^{*} (n_0))x_{\sigma b}$, where we set $U_{\Lambda } = -28 \,{\rm MeV}$, $U_{\Sigma } = 30 \,{\rm MeV}$, and $U_{\Xi } = -15 \,{\rm MeV}$.
The $\Delta$ potential is poorly constrained by the data. We explore the range $-50\,\mev < U_\Delta < -100\,\mev$, keeping in mind that the realistic value of $U_\Delta$ is close to the nucleon one $U_N \sim -(50\mbox{--}60)\,\mev$. The scaling function for $\phi$ meson is chosen as $\eta_\phi = (1-f)^2$, which corresponds to the $H\phi$ set of models in~\cite{Maslov:2015wba}. Below we label the MKVOR* model with $\Delta$s and without hyperons as MKVOR*$\Delta$, and the model with $\Delta$s and full baryon octet with $H\phi$ scaling as MKVOR*H$\Delta\phi$. Neutron star masses and radii are obtained by integrating the Tolman--Oppenheimer--Volkoff equation. For the BEM for densities $n\leq0.75n_0$ our EoS is supplemented by the EoS of the NS crust. The details  can be found in~\cite{Maslov:2015wba}.

\begin{figure}
\centering
\includegraphics[width=.65\textwidth]{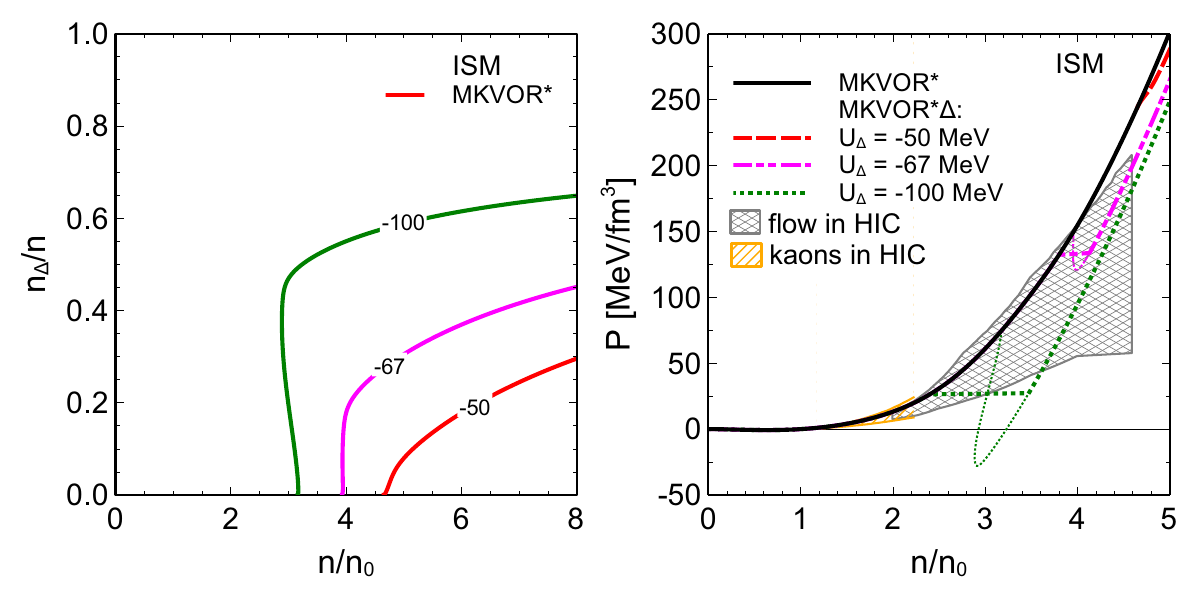}
\caption{Left panel: $\Delta$ concentrations in ISM as functions of the density. Lines are labeled with the potentials $U_\Delta$, in $\mev$.   Right panel: Pressure as a function of the density for our model without and with $\Delta$s for $U_\Delta=-50\,\mev$, $U_\Delta=-67\,\mev$ and $U_\Delta=-100\,\mev$. For $U_\Delta=-67\,\mev$ and $-100\, \mev$ bold lines show the Maxwell constructions and thin lines show the original pressure curves. Hatched area shows the flow constraint from \cite{Danielewicz:2002pu}.}
\label{dsym}
\end{figure}

\section{Numerical results}
On the left panel in figure~\ref{dsym} we show the $\Delta$ concentrations, determined in the ISM by the chemical equilibrium condition $\mu_N(n, n_\Delta) = \mu_\Delta(n,n_\Delta)$. We see that with a decrease of $U_\Delta$ the critical density for the $\Delta$ appearance decreases. For $U_\Delta > -55 \,\mev$ the $\Delta$ baryons appear in a third-order phase transition. Interestingly, for $U_\Delta < -67\,\mev$  there exists a density range with multiple solutions for $n_\Delta$ at a given density. Among these solutions the one with greater $n_\Delta$ is energetically favorable. This means that $\Delta$s appear with a jump from $n_\Delta = 0$ to a finite value of $n_\Delta$ in a first-order phase transition.
The pressure in the ISM as a function of the density is shown on the right panel in figure~\ref{dsym}.  For $U_\Delta < -55\,\mev$ in the equilibrium the system follows the Maxwell-construction line. For $U_\Delta < -67\,\mev$ the region with multiple solutions manifests itself also in the pressure. Note that for $-83\,\mev < U < -65\,\mev$ the equilibrium pressure curve lies fully within the flow constraint. It means that if   in the future the constraint sharpens, it could be used to constrain the $\Delta$ potential.
\begin{figure}
\centering	
\includegraphics[width=.65\textwidth]{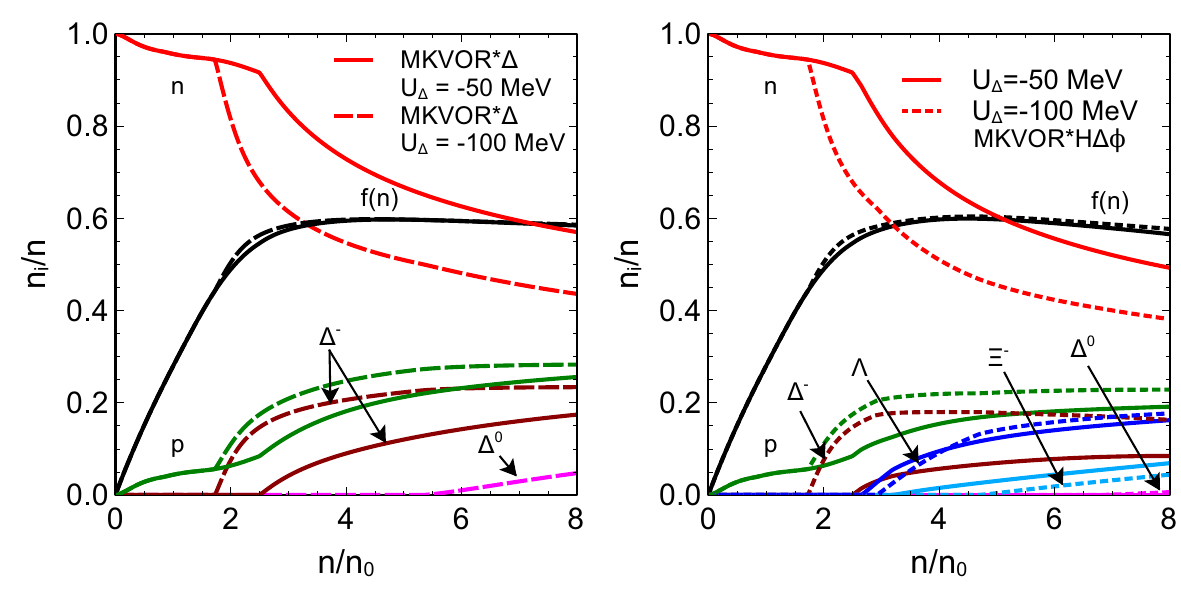}
\caption{Particle fractions and the scalar field $f$ as functions of the density in BEM in the MKVOR*$\Delta$ (left panel) and MKVOR*H$\Delta\phi$ (right panel) models for $U_\Delta = -50\,\mev$ and $-100\,\mev$.}
\label{nd_bem}
\end{figure}

The particle fractions as functions of the density in the BEM are presented in figure~\ref{nd_bem} for the MKVOR*$\Delta$ model (left panel) and for the MKVOR*H$\Delta\phi$ model (right panel), for $U_\Delta = -50\,\mev$ and $ -100\,\mev$. With an increase of the density $\Delta^-$s appear  in both cases at density $n = 2.51 \, n_0$ for $U_\Delta = -50\,\mev$ and $n=1.74 \,n_0$ for $U_\Delta = -100\,\mev$. The $\Delta$ fraction increases sharply, and for the MKVOR*$\Delta$ model reaches $0.23$ and $0.17$ for $U_\Delta=-50\, \mev$ and $-100\, \mev$, respectively. As in \cite{Drago2014}, the inclusion of hyperons reduces the amount of $\Delta$s at a given density.

\begin{figure}
\centering
\includegraphics[width=\textwidth]{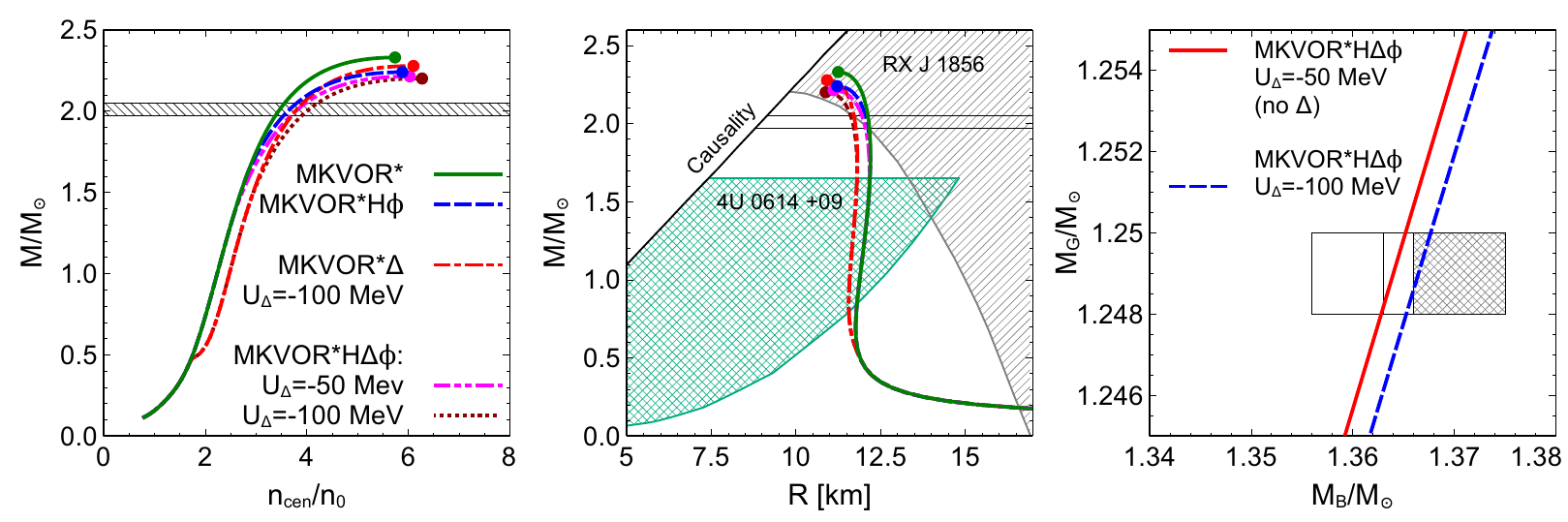}
\caption{The NS mass versus the central density (left panel) and NS radius (middle panel) in the MKVOR*, MKVOR*H$\phi$, MKVOR*$\Delta$ ($U_\Delta = -100\,\mev$) and MKVOR*H$\Delta\phi$ ($U_\Delta = -50\,\mev $ and $-100\, \mev$ )  models. Right panel: The gravitational NS mass as a function of the baryon mass for MKVOR*H$\Delta\phi$ model  for $U_\Delta = -50\,\mev$ ($\Delta$s do not yet appear for stars with masses in the shown range) and $U_\Delta = -100\,\mev$. Shaded rectangle shows the constraint from~\cite{Podsiadlowski}, two empty rectangles show the change of the constraint at the assumption of progenitor mass loss by $0.3\%M_\odot$ and $1\%M_\odot$.}
\label{NSprops}
\end{figure}

Despite the $\Delta$ fraction in the NS medium is large, the effect of $\Delta$s on the NS masses and radii appears to be small. On the left panel in figure~\ref{NSprops} we show the NS mass as a function of the central density for our models,  and on the middle panel the mass-radius plot is shown. The maximum decrease of a NS mass at a given central density in MKVOR*H$\Delta\phi$ model is less than $0.02 \, M_\odot$ for $U_\Delta = -50\,\mev$ and does not exceed $0.2 \, M_\odot$ for $U_\Delta = -100\,\mev$.   The decrease of the maximum NS mass is even smaller, being less than $0.05 \, M_\odot$ for $U_\Delta = -100\,\mev$. The  change of the NS radius for a given NS mass is not more than $0.5 \, {\rm km}$. Thus, within the MKVOR*H$\Delta\phi$ model both $\Delta$ and hyperon puzzles are resolved.

An interesting observation is that the inclusion of $\Delta$s allows to improve the constraint extracted from analysis of gravitational and baryon masses of the pulsar J0737-3039(B)~\cite{Podsiadlowski}. On the right panel in figure~\ref{NSprops} we show the gravitational NS mass as a function of the baryon NS mass for $U_\Delta = -50\,\mev$ and $-100\,\mev$ together with the constraint from~\cite{Podsiadlowski}. For $U_\Delta = -50\,\mev$ the curve is the same as in the case without $\Delta$s, since $\Delta$s do not appear at the densities corresponding to the central density of the NS with $M_G \simeq 1.25 \, M_\odot$. For $U_\Delta = -100\,\mev$ the constraint is passed better.

\section{Conclusion}
We studied the effect of inclusion of $\Delta$ isobars on the EoS of the NS matter with hyperons within the  MKVOR-based RMF models of~\cite{Maslov:2015msa,Maslov:2015wba} with scaled hadron coupling constants and masses. We varied the $\Delta$ potential at the saturation density, $U_\Delta$, in the range $-100\,\mev < U_\Delta < -50\,\mev$ to estimate the maximum effect of the $\Delta$ appearance on NS properties. For the ISM $\Delta$ baryons can appear in third- or first-order phase transitions, depending on the value of $U_\Delta$. In the NS matter the $\Delta$ fraction is not small for $-100\,\mev < U_\Delta < -50\,\mev$,  which we studied. In spite of that, a decrease of the maximum NS mass does not exceed $0.05 M_\odot$, so the maximum mass constraint remains satisfied.
In the presence of $\Delta$s the constraint on the relation between the gravitational and baryon masses of J0737-3039(B)~\cite{Podsiadlowski} proves to be better fulfilled.

\ack
The reported study was funded by the Russian Foundation for Basic Research (RFBR) according to the research project No 16-02-00023-A. The work was also supported by the Slovak Grant No. VEGA-1/0469/15, by ``NewCompStar'', COST Action MP1304 and by the Ministry of Education and Science of the Russian Federation (Basic part).


\end{document}